\documentclass[aps,pre,twocolumn,showpacs]{revtex4}
\usepackage{epsfig}
\usepackage{times}
\begin{document}

\title{Ordering in spatial evolutionary games for pairwise collective strategy updates}

\author{Gy\"orgy Szab\'o,$^{1}$ Attila Szolnoki,$^{1}$ Melinda Varga,$^{2}$ and L\'{\i}via Hanusovszky$^{3}$}
\affiliation {
$^{1}$Research Institute for Technical Physics and
Materials Science, P.O. Box 49, H-1525 Budapest, Hungary \\
$^{2}$ Babes-Bolyai University, Faculty of Physics, RO-300223 Cluj-Napoca, Romania \\
$^{3}$ Roland E{\"o}tv{\"o}s University, Institute of Physics, P\'azm\'any P. s\'et\'any 1/A, H-1117 Budapest, Hungary}

\begin{abstract}
Evolutionary $2 \times 2$ games are studied with players located 
on a square lattice. During the evolution the randomly chosen neighboring players try to maximize their collective income by adopting a random strategy pair with a probability dependent on the difference of their summed payoffs between the final and initial state assuming quenched strategies in their neighborhood. In the case of the anti-coordination game this system behaves alike an anti-ferromagnetic kinetic Ising model. Within a wide region of social dilemmas this dynamical rule supports the formation of similar spatial arrangement of the cooperators and defectors ensuring the optimum total payoff if the temptation to choose defection exceeds a threshold value dependent on the sucker's payoff. The comparison of the results with those achieved for pairwise imitation and myopic strategy updates has indicated the relevant advantage of pairwise collective strategy update in the maintenance of cooperation.
\end{abstract}


\maketitle

\section{Introduction}
\label{sec:introduction}

Most of the games represent simplified real-life situations and help us to find an optimum decision (action). Due to the simplifications the players have only a few options to choose and the corresponding incomes are quantified by a payoff matrix allowing us to apply the tools of mathematics. The theory of games has been used successfully both in economics and political decisions since the pioneering work of von Neumann and Morgenstern \cite{neumann_44}. Subsequently the concept of payoff matrix is adopted by biologists to quantify the effect of interactions of species on their fitness (characterizing the capability to create offspring) in the mathematical models of Darwinian evolution \cite{maynard_82}. Since that time the evolutionary game theory provides a general mathematical framework for the investigation of multi-agent systems used widely in economy and other social sciences where imitation is substituted for the offspring creation  \cite{samuelson_97,hofbauer_98,nowak_06}. 

In traditional game theory selfish and intelligent individuals try to maximize their own payoff irrespective of others. In evolutionary game theory players repeat the games and sometimes imitate (adopt) the neighbor's strategy if the neighbor received a higher score. It turned out that assuming local interactions among players the imitation supports the maintenance of altruistic behavior even for Prisoner's Dilemma games where the individual interest is in conflict with the common one  and the selfish individual behavior drives the well-mixed community into a state (called "tragedy of the commons") with players exploiting (instead of helping) each other \cite{axelrod_84,nowak_s06,nowak_06,szabo_pr07,sigmund_10}. 

In parallel with theoretical investigations game theory is also used to study human and animal behaviors experimentally \cite{flood_ms58,milinski_prslb97,fischbacher_el01,fehr_n02,
henrich_bbs05,fehr_n08,clutton_brock_n09,tricomi_n10}. These experiments have motivated the extension of evolutionary games to study the effect of different types of mutual help, e.g., charity \cite{li_pa10}, inequality (inequity) aversion \cite{fehr_qje99,xianyu_pa10,scheuring_jtb10}, emotions \cite{kirman_ptrsb10} including juvenile-adult interactions \cite{lion_tpb09}. In generally, the modelling of human decision dynamics is one of the most important open problems in the behavioral sciences \cite{traulsen_pnas10}. The examples mentioned raise the possibility that a player tries to optimize not only personal but her local neighborhood's payoff as well. Motivated by this option we consider the simplest case and introduce a collective pairwise strategy update rule providing that two randomly chosen neighbors upgrade their strategy simultaneously in order to increase their summarized payoff each coming from games with all their neighbors on the spatial system. This way of strategy update can be considered as an extension of cooperative games toward the spatial evolutionary games. Originally, in cooperative games groups of players (coalitions) may perform coordinated behavior within the group to enhance the group's payoff. On the other hand, the present model implies a connection between the theory of kin selection \cite{hamilton_jtb64a,hamilton_jtb64b} and spatial models of viscous population of altruistic relatives helping each other \cite{vanbaalen_jtb98,doebeli_el05,grimm_geb09}. 

As a consequence of the proposed strategy update rule, it will be shown that the previously mentioned ''tragedy of the commons'' state can be avoided even in the hard condition of Prisoner's Dilemma game. In the latter case both analytical and numerical approaches indicate the existence of an ordered structure of cooperator and defector players on square lattice at sufficiently low noise level. (This arrangement of alternative strategies resembles the sublattice ordering of anti-ferromagnetic Ising model.) It is worth mentioning that similar formation of strategies was also reported by Bonabeau {\it et al.} \cite{bonabeau_pa95} and by Weisbuch and Stauffer \cite{weisbuch_pa07} within the framework of social models. To explore and identify the exclusive consequence of the proposed strategy update rule, we will compare the results with the outcomes of two previously applied dynamical rules. These are the imitation of a better neighbor and the so-called myopic strategy update rules.

The present work is structured as follows. In Sec.~\ref{sec:models} we define the spatial evolutionary games with the mentioned dynamical rules. The main results for these types of dynamical rules are compared for the anti-coordination game in Sec.~\ref{sec:ACG}. Subsequently we will discuss the weak Prisoner's Dilemma games with using Monte Carlo (MC) simulations and mean-field analysis. In Sec.~\ref{sec:robus} we present and compare the MC results for the three dynamical rules within a relevant region of payoff parameters describing social dilemmas \cite{dawes_arp80}. As the dynamical rules influence significantly the sublattice ordering process therefore some aspects of domain growth are considered numerically in Sec.~\ref{sec:domgr}. Finally we summarize the main results in Sec.~\ref{sec:conc}.  

\section{Models with different dynamical rules}
\label{sec:models}

In the studied models each player follows one of the pure cooperate or defect ($C$ or $D$) strategies. According to pairwise interaction a player's payoff is calculated by means of $2~\times~2$ payoff matrix. For a given pair of equivalent players the possible strategy dependent payoffs are given by the payoff matrix as
\begin{equation}
{\bf A}=\left( \matrix{a & b \cr
                       c & d \cr} \right)\;, 
\label{eq:pom}
\end{equation} 
where $a$ ($d$) is received by the $C$ ($D$) player if her co-player follows the same strategy. On the other hand, if the players choose opposite strategies the $C$ players receive $b$ while $D$s are rewarded by $c$. The anti-coordination (AC) game will be considered when $a=d=0$ and $b=c=1$. In this case the players receive the maximum payoff if they choose opposite strategies. For the social dilemmas we also use a rescaled payoff matrix \cite{santos_pnas06} in such a way that $a=1$ and $d=0$, that is, the mutual choice of $C$ is better for both players. Despite it the players can favor the choice of $D$ if either $c=T>1$ or $b=S<0$ where $T$ refers to the temptation to choose defection and $S$ is the sucker's payoff. For the Prisoner's Dilemma (PD) both conditions are satisfied and the players are enforced to choose $D$ yielding the second lowest individual income for them. The Hawk-Dove, in short HD (also called as Snowdrift or Chicken) game describes the situation when $T>1$ and $S>0$ while the Stag-Hunt (SH) game corresponds to the case $T<1$ and $S<0$. The fourth quadrant of the $T-S$ parameter plane is represented by the Harmony (H) game where mutual $C$ is the best solution for the players. In the mentioned four quadrants of the $T-S$ plane the two-person one-shoot games have different set of Nash equilibria \cite{hofbauer_98,gintis_00,nowak_06,szabo_pr07}.

In the present spatial models players are located on the sites $x$ of a square lattice consisting of $L \times L$ nodes under periodic boundary conditions. Initially each player follows an $s_x=C$ or $D$ strategy chosen at random. The payoff $P_x$ is collected from the mentioned matrix games with her four nearest neighbors. According to the proposed pairwise collective strategy update the evolution of strategy distribution is based on the following protocol. First, we choose two neighboring players ($x$ and $y$) at random and we evaluate their payoff ($P_x$ and $P_y$) depending on their own $s_x$, $s_y$, and also on the neighboring strategies. Subsequently we evaluate the payoff $P_x^{\prime}$ and $P_y^{\prime}$ assuming that the given players follow randomly chosen $s_x^{\prime}$ and $s_y^{\prime}$ strategies while the neighborhood remains unchanged. As a consequence of randomly chosen $s_x^{\prime}$, $s_y^{\prime}$ strategy pair, there are cases when only one (or none) of the two players will modify her strategy. Notice, however, that this strategy choice allows the pair of players to select all the possible four strategy pairs. Finally the strategy pair, $s_x^{\prime}$ and $s_y^{\prime}$, is accepted simultaneously with a probability 
\begin{equation}
W_c =\frac{1}{1+\exp[(P_x+P_y-P_x^{\prime}-P_y^{\prime})/K]}\;,
\label{eq:pwc}
\end{equation}
where $K$ characterizes the average amplitude of noise disturbing the players' rational decision. 

The results of the above evolutionary process will be contrasted with the consequence of two other dynamical rules used frequently in previous studies \cite{nowak_06,szabo_pr07}. If the evolution is controlled by stochastic imitation of the more successful neighbor then player $x$ adopts the neighboring strategy $s_y$ with a probability
\begin{equation}
W_i =\frac{1}{1+\exp[(P_x-P_y)/K]}\
\label{eq:imit}
\end{equation}
dependent on the current payoff difference between players $x$ and $y$. Besides it, we also study a so-called myopic strategy update when a randomly chosen player $x$ changes her strategy $s_x$ to a random strategy $s_x^{\prime}$ with a probability
\begin{equation}
W_m =\frac{1}{1+\exp[(P_x-P_x^{\prime})/K]}\ \,
\label{eq:myop}
\end{equation}
where $P_x$ and $P_x^{\prime}$ are the income of player $x$ when playing $s_x$ and $s_x^{\prime}$ for the given neighborhood. Notice that the latter strategy update is analogous to the Glauber dynamics used in the kinetic Ising models \cite{glauber_jmp63}. Consequently, for symmetric payoff matrices, $b=c$, (or potential games) the myopic strategy update drives the spatial system into a thermal equilibrium (at temperature $K$) that can be described by the Boltzmann statistics \cite{blume_geb03,szabo_pr07}. This means that an anti-ferromagnetic ordering process is expected for the AC games with myopic strategy update when decreasing the noise parameter $K$.

\section{Results for anti-coordination game}
\label{sec:ACG}

Motivated by the above mentioned connection to the Ising model, we first consider the anti-coordination game and study the consequences of different strategy update rules. The presented results of MC simulations were obtained typically for $L=400$ size but we used significantly larger system size in the vicinity of the critical transitions to suppress undesired fluctuations. During the evolution we have determined the average portion $\rho$ of players following the $C$ strategy in the stationary state. To describe the expected anti-ferromagnetic ordering the square lattice is divided into two sublattices ($A$ and $B$) on the analogy of white and black boxes on the chessboard. In fact two equivalent types of completely ordered structure exist in the limit $K \to 0$. For both cases the $C$ and $D$ strategies are present with the same frequency ($\rho=1/2$). In the first (second) case all the $C$ strategies are located on the sites of sublattice $A$ ($B$). The sublattice ordering will be characterized by an order parameter $M=|\rho_A-\rho_B|$ where $\rho_A$ and $\rho_B$ denote the portion of $C$ strategy in the sublattices $A$ and $B$. In a finite system the sublattice ordering develops throughout a domain growing process within a transient time.

\begin{figure}[ht]
\centerline{\epsfig{file=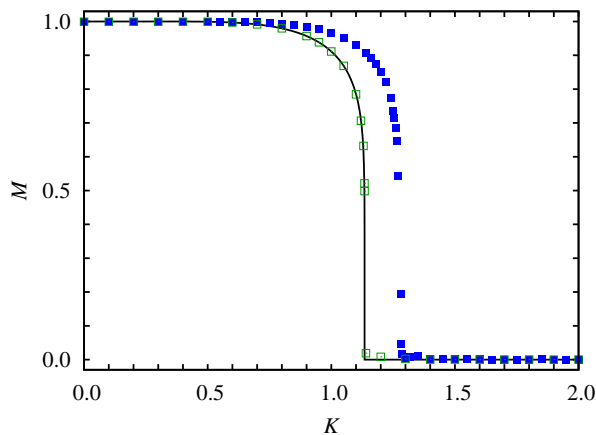,width=8cm}}
\caption{\label{fig:ising}(color online) Order parameter $M$ as a function of $K$ in the anti-coordination games for myopic (green open squares) and pairwise collective strategy updates (blue closed squares). The solid line shows the theoretical result obtained by Onsager \cite{onsager_pr44} for the two-dimensional Ising model.}
\end{figure}

Starting with myopic strategy update, defined by Eq.~\ref{eq:myop}, the MC data coincide with the exact results of Ising model \cite{onsager_pr44} if the noise parameter (temperature in the latter case) is rescaled by a factor of 2. Accordingly, a long-range ordered state appears in the zero noise limit where cooperator and defector players form a chessboard-like pattern. The order parameter $M$ varies from 1 to 0 if $K$ is increased from 0 to $K_c=1/\ln(1+\sqrt{2})$ and $M=0$ if $K>K_c$, as illustrated in Fig.~\ref{fig:ising}.
When considering the analogy between the kinetic Ising spin systems and evolutionary AC games one should keep in mind that the Glauber dynamics \cite{glauber_jmp63} favors spin flips decreasing the total energy and the opposite flips are generated by the external noise (temperature). On the contrary, for the evolutionary games the myopic individuals wish to increase their own payoff and the opposite decision is caused by noisy effects. The necessity of the temperature rescaling is related to the fact that in the kinetic Ising model for Glauber dynamics the individual spin flips are controlled by the total energy difference while for the evolutionary games the changes are influenced by the individual payoff increase ($\Delta P_x$) that is half of the total payoff increase (if the payoff matrix is symmetric) because the co-players share the income equally.

In the following we study the evolutionary AC game with pairwise collective strategy update rule defined by Eq. (\ref{eq:pwc}). In agreement with our expectation this strategy update is capable to find the optimal global state and the long-range sublattice ordering is established again when varying the amplitude of noise.
Figure~\ref{fig:ising} indicates a striking qualitative similarity between the behaviors controlled by the myopic and pairwise collective strategy updates. The slight difference between the outputs is related to the fact how payoffs change due to different update processes. As we already mentioned, the change of individual payoff is half of the global change for myopic rule. This is not necessarily true for pairwise collective rule because neighboring players receive nothing from the game played between the focal $x$ and $y$ players. In a special case, when focal players adopt each others strategies simultaneously, the exact relation between the changes of individual and global payoffs is restored. This special process, when $s_x^{\prime}=s_y$ and $s_y^{\prime}=s_x$, resembles the Kawasaki spin-exchange of statistical mechanics. As a consequence, this ``limited'' pairwise collective dynamics reproduces the results of myopic update.

To close this section, we should stress that the application of the third type pairwise imitation strategy update is unable to find the optimal solution when cooperator and defector players form a long-range ordered state. In the following we will study the consequences of different updating rules when the payoff matrix is not symmetric.

\section{Results for weak prisoner's dilemma}
\label{sec:wPD}

We start with a simple and popular parametrization of the weak PD game when ($S=-0$). If we fix the noise level, the only free parameter is the $T$ temptation to defect payoff element. In the rest of the paper we use $K=0.25$ noise value, that is proved to offer an almost optimal cooperation level in case of pairwise imitation strategy update model \cite{szabo_pre05}.

\subsection{Mean-field theory}

The pairwise imitation strategy update rule was already investigated by means of mean-field theory
(for a brief survey see \cite{nowak_06,szabo_pr07} and further references therein). Within this approach the stationary state is characterized by the average fraction $\rho$ of cooperators that drops suddenly (at $T=1$) from 1 to 0 if $T$ is increased for arbitrary values of $K$. For completeness, we note that the results for arbitrary values of $S$ are surveyed in a recent review \cite{roca_plr09} using replicator dynamics in the imitation of the better strategies, too. 

As expected, the application of the other two strategy update rules, such as myopic and pairwise collective, may allow the possibility of sublattice ordering. To catch this behavior we extend the mean-field analysis and introduce two sublattices ($A$ and $B$) where the fraction of cooperators can be different ($\rho_A$ and $\rho_B$, respectively). Using the general payoff parameters given by Eq. (\ref{eq:pom}) the average payoff of cooperator and defector players in the sublattices $A$ and $B$ can be approximated as
\begin{eqnarray}
P_A^{(C)} &=& 4 [a \rho_B + b (1-\rho_B)], \nonumber \\
P_A^{(D)} &=& 4 [c \rho_B + d (1-\rho_B)], \nonumber \\ 
P_B^{(C)} &=& 4 [a \rho_A + b (1-\rho_A)], \nonumber \\ 
P_B^{(D)} &=& 4 [c \rho_A + d (1-\rho_A)], \;
\label{eq:Us}
\end{eqnarray}
for the present connectivity structure where each player in sublattice $A$ has four neighboring players belonging to sublattice $B$ and vice versa.

For the myopic strategy update [defined by Eq.~(\ref{eq:myop})] the time derivative of cooperator frequency $\rho_A$ can be expressed as:
\begin{eqnarray}
\dot{\rho}_A = &-& \rho_A \frac{1}{1+\exp[(P_A^{(C)}-P_A^{(D)})/K]} \nonumber \\
&+& (1-\rho_A) \frac{1}{1+\exp[(P_A^{(D)}-P_A^{(C)})/K]} \;, 
\label{eq:rhoevol}
\end{eqnarray}
and a similar expression can be derived for $\dot{\rho}_B$ by substituting $B$ for the sublattice index $A$. Inserting the expressions (\ref{eq:Us}) into (\ref{eq:rhoevol}) after some mathematical manipulations one finds
 \begin{eqnarray}
\dot{\rho}_A &=& \frac{1}{1+\exp \{ 4 [(b+c-a-d)\rho_B-b+d]/K \} } - \rho_A \;, \nonumber \\
\label{eq:rhoabeq}\\
\dot{\rho}_B &=& \frac{1}{1+\exp \{ 4 [(b+c-a-d)\rho_A-b+d]/K \} } - \rho_B \nonumber \;. 
\end{eqnarray}

The stationary values of cooperator densities $\rho_A$ and $\rho_B$ can be determined numerically from Eq. (\ref{eq:rhoabeq}) when $\dot{\rho}_A=\dot{\rho}_B=0$. For the case of the weak PD ($a=1$, $b=S=0$, $c=T$, and $d=0$) the $T$-dependence of the stationary solution exhibits two types of behaviors as illustrated by green lines in the upper plot of Fig.~\ref{fig:rho_wPD}. Below a threshold value (depending on $K$) the distribution of cooperators is homogeneous, that is, $\rho_A=\rho_B$. Above the threshold temptation value the mean-field solution predicts a (twofold degenerated) sublattice ordering. For one of the ordered structure (at sufficiently high values of $T$) sublattice $A$ is occupied dominantly by defectors and in sublattice $B$ the players alternate their strategies between cooperation and defection (i.e., $\rho_B \simeq 0.5$) because both strategies yield the same payoff for them (notice that $P_B^{(C)}=P_B^{(D)}=0$ if $\rho_A=0$). For the second (equivalent) ordered structure the role of sublattices $A$ and $B$ is exchanged. We should mention that there is an unstable symmetric solution without sublattice ordering ($\rho_A=\rho_B$) for high $T$ values as illustrated by spaced dashed green line in the upper plot of Fig.~\ref{fig:rho_wPD}.

The mean-field analysis can also be performed for the pairwise collective strategy update although the corresponding formulae become more complicated because of the larger number of elementary events (single and simultaneous two-site strategy flips for two neighboring players). Neglecting the technical details now we only present the numerical results by blue lines in the upper plot of Fig.~\ref{fig:rho_wPD}. Similarly to the myopic case, the stable solution is a homogeneous state for low $T$ values and a sublattice ordering appears above a threshold temptation. In the latter case, one of the sublattices is occupied dominantly by cooperators while the other sublattice is occupied by defectors. Naturally, this solution is twofold degenerate if we substitute the role of sublattices $A$ and $B$. An unstable symmetric solution also exists at high $T$ values that is marked by spaced dotted blue line in the upper plot of Fig.~\ref{fig:rho_wPD}.

In the following we will check the predictions of mean-field theory by using MC simulations.

\subsection{Simulations}

\begin{figure}[ht]
\centerline{\epsfig{file=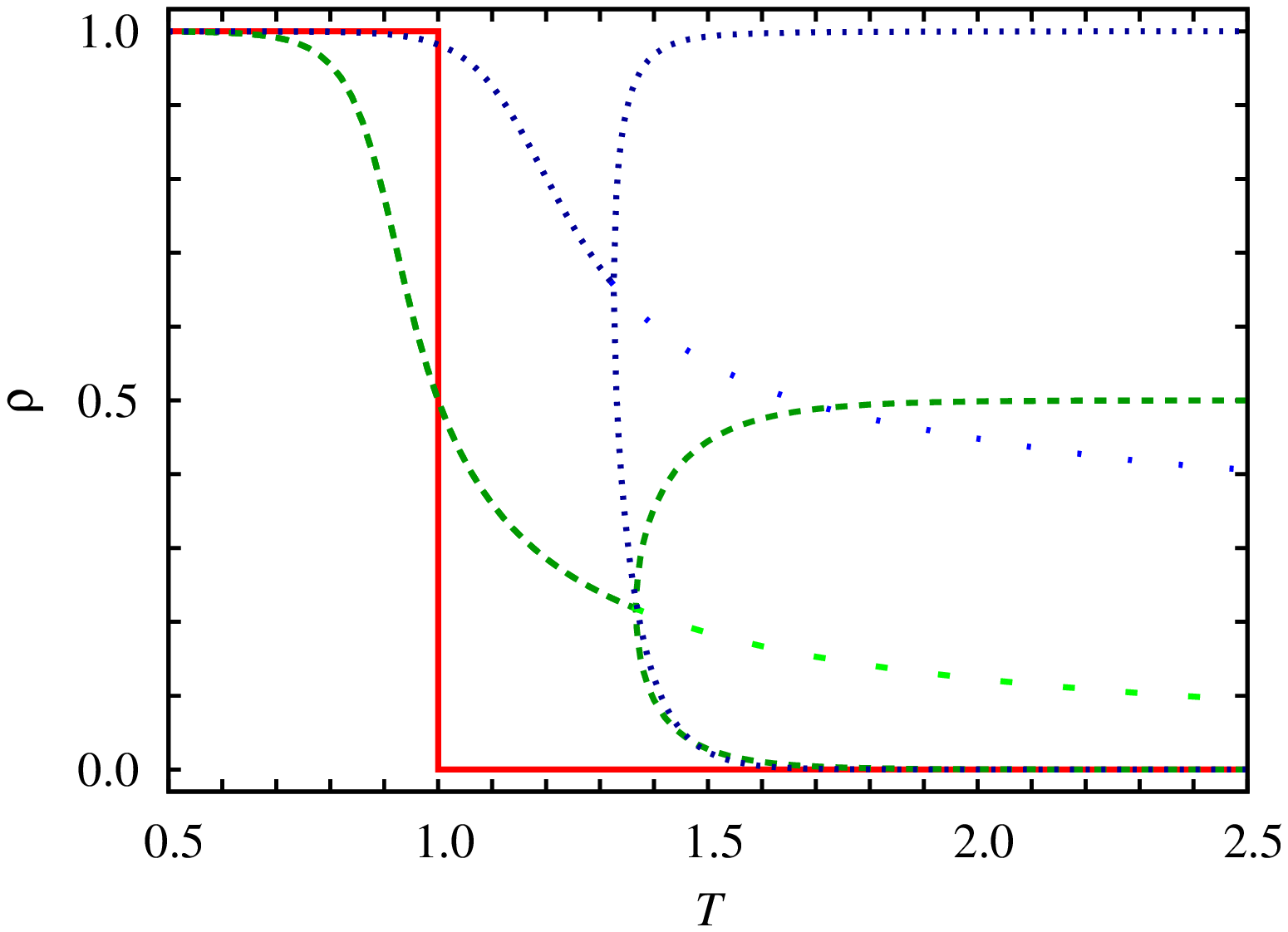,width=8cm}}
\centerline{\epsfig{file=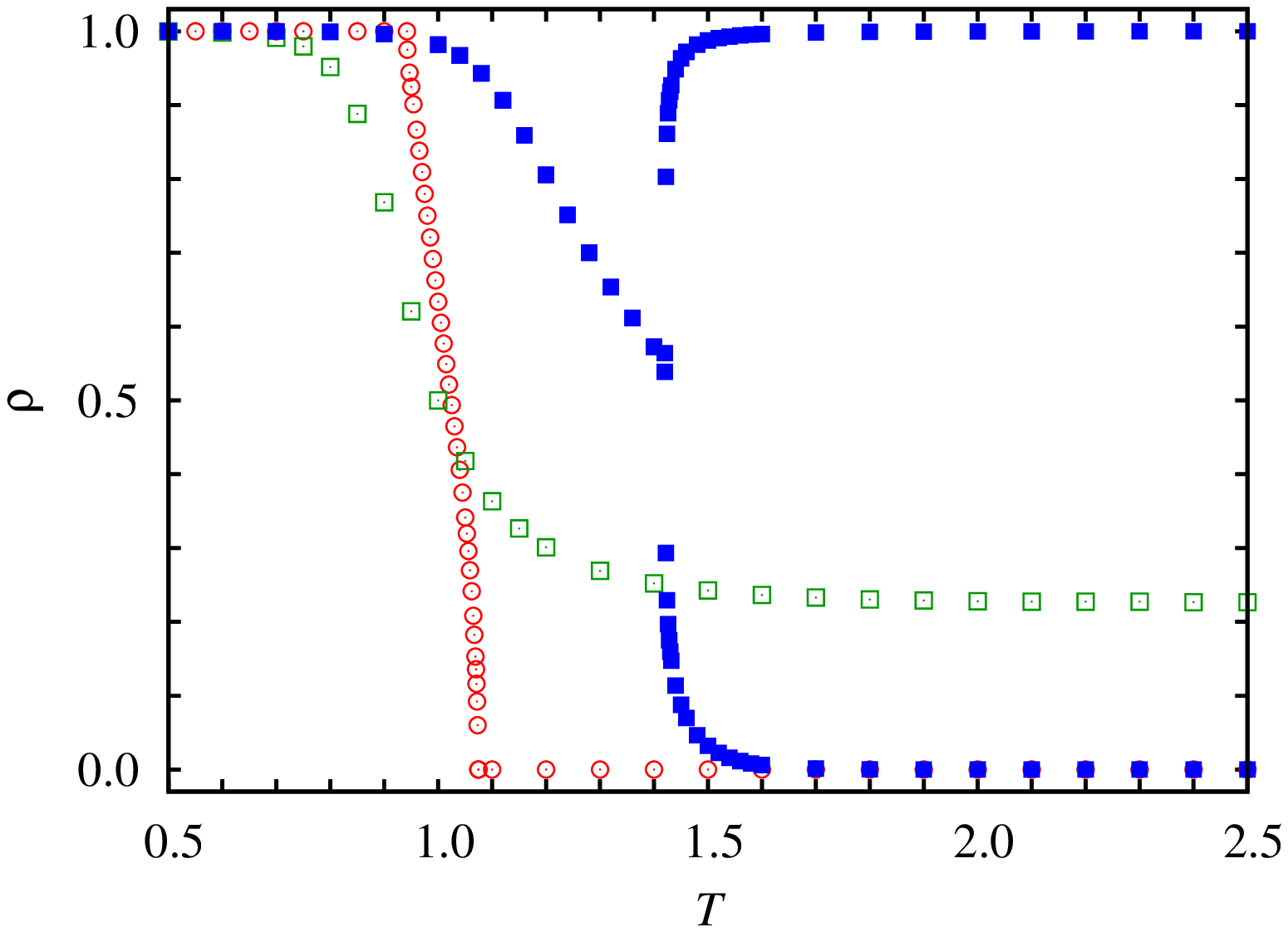,width=8cm}}
\caption{\label{fig:rho_wPD}(color online) Density of cooperators in the sublattices as a function of $T$ for the weak PD game on a square lattice at $K=0.25$. The upper plot shows solutions predicted by mean-field theory. Solid (red), dashed (green), and dotted (blue) lines illustrate the stable solutions for pairwise imitation, myopic, and pairwise collective strategy update rules, respectively. Unstable solutions of the mean-field equations at high $T$ values are denoted by spaced dashed and dotted lines for the last two dynamical rules. Lower plot shows the corresponding MC results. Using the same color coding, red open circles are for pairwise imitation, green open squares for myopic, and blue closed squares are for pairwise collective rules.}
\end{figure}

Similarly to the AC model, the typical system size was $L=400$ during the simulations. In the vicinity of phase transition point, however, we had to use larger systems (up to $L=2000$) to gain the sufficient accuracy. The results for different strategy update are summarized in the lower panel of Fig.~\ref{fig:rho_wPD}. For completeness, we quote here the known MC results of pairwise imitation update \cite{szabo_pre05}. These results are marked by red open circles in the plot. In contrast to mean-field prediction, $C$ and $D$ strategies can coexist if $T_{c1}<T<T_{c2}$ (where $T_{c1}=0.942(1)$ and $T_{c2}=1.074(1)$ for $K=0.25$). For lower values of $T$ only the $C$ strategy can remain alive in the final stationary state while $C$ becomes extinct if $T>T_{c2}$.   

In case of myopic strategy update, qualitatively similar behavior was found (marked by green open squares in the lower plot of Fig.~\ref{fig:rho_wPD}). The only difference is a relatively high portion of cooperator players in the high $T$ region. The survival of the $C$ strategy is caused by appearance of solitary $C$s in the sea of $D$s because this event does not modify the payoff of the given individual if $S=0$. Evidently, the probability of the mentioned process decreases if $S$ becomes negative, as it happens in the real PD situations. 

In contrast to the prediction of mean-field calculation the MC simulations do not justify the presence of long-range ordering. Instead of it the simulations give $\rho_A=\rho_B \simeq 0.2265(2)$ if $T$ is sufficiently high. It is worth mentioning that the more sophisticated pair approximation (the results are not indicated in the figure) reproduces the absence of sublattice ordering in the case of $S=0$. For slightly higher values of $S$ (in the region of HD game), however, the prediction of mean-field calculation becomes qualitatively correct because MC simulations indicate sublattice ordering as detailed later on.

Closing by pairwise collective update, MC data (blue closed squares) fully support the prediction of mean-field calculations and show a similar sublattice ordering as we observed for AC game previously. Namely, the portion of $C$ strategy is distinguishable in the sublattices $A$ and $B$, if $T>T_c(K=0.25)=1.41(1)$. Notice furthermore, that the average density of $C$s is significantly higher for this strategy update in comparison with those provided by the pairwise imitation and myopic rules. 

\begin{figure}[ht]
\centerline{\epsfig{file=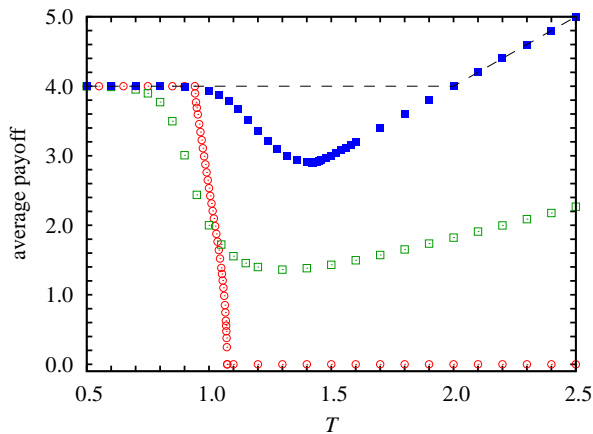,width=8cm}}
\caption{\label{fig:apo}(color online) Average payoff versus $T$ for three dynamical rules at $K=0.25$ in the case of weak PD. The MC results are illustrated by red open circles for pairwise imitation, green open squares for myopic and blue closed squares for pairwise collective strategy adoption rule. The dashed line denotes the maximum average payoff available in the system.}
\end{figure}

The pairwise collective strategy update promotes the chessboard-like arrangement of cooperators and defectors because this constellation provides the highest total income for the neighboring co-players (as well as for the whole community) if $T+S>2R$, where $2R$ is the total payoff of a cooperator pair. This feature can be clearly recognized for the weak PD game where the ordered structure is disturbed rarely by point defects if $T>2$ and the average payoff increases linearly with $T$ as illustrated in Fig.~\ref{fig:apo}. It is well known that the traditional game theory \cite{sigmund_10} suggests the players to alternate $C$ and $D$ in opposite phase to receive $(T+S)/2$ on average for the repeated two-person games. The sublattice ordering in the spatial evolutionary game can be considered as an alternative solution to achieve the maximum average payoff. 

From the view point of the pair of players (or the whole society) the chessboard-like arrangement of cooperation and defection has an advantage over the state of the homogeneous defection favored by several dynamical rules within a wide payoff region of PD. As a result, the ordered structure is observable in the spatial strategy distribution even for $T<2$ for the weak PD. Below a threshold value of $T$, however, the long-range ordered strategy distribution is destroyed by the noise and both types of ordered structures are present within small domains. Simultaneously, the further decrease of $T$ increases the frequency of cooperators approaching to the saturation value $\rho=1$. The quantitative analysis of the pair's payoff ($P_x+P_y$) shows that the homogeneous cooperation remains stable against the appearance of a solitary $D$ if $T<5/4$ for the case of weak PD. Similarly, in the perfect sublattice ordered state ($\rho_A=1$ and $\rho_B=0$) the appearance of an additional cooperator is preferred if $T<5/4$. At the same time the homogeneous defection is not stable because any new cooperator increases the income of her neighbors (and also the income of pairs she belongs to). From the above features one can conclude a sharp transition at $T=5/4$ from the homogeneous $C$ state ($\rho_A=\rho_B=1$) to one of the chessboard-like structure (e.g., $\rho_A=1$ and $\rho_B=0$) if $T$ is increased in the limit $K \to 0$. This expectation is confirmed by MC simulations performed for several low values of noise $K$. 

In the following we extend the payoff parametrization to general social dilemmas and explore how robust the observed long-range ordering in the whole $T-S$ plane.

\section{Results for social dilemmas}
\label{sec:robus}

\begin{figure}[ht]
\centerline{\epsfig{file=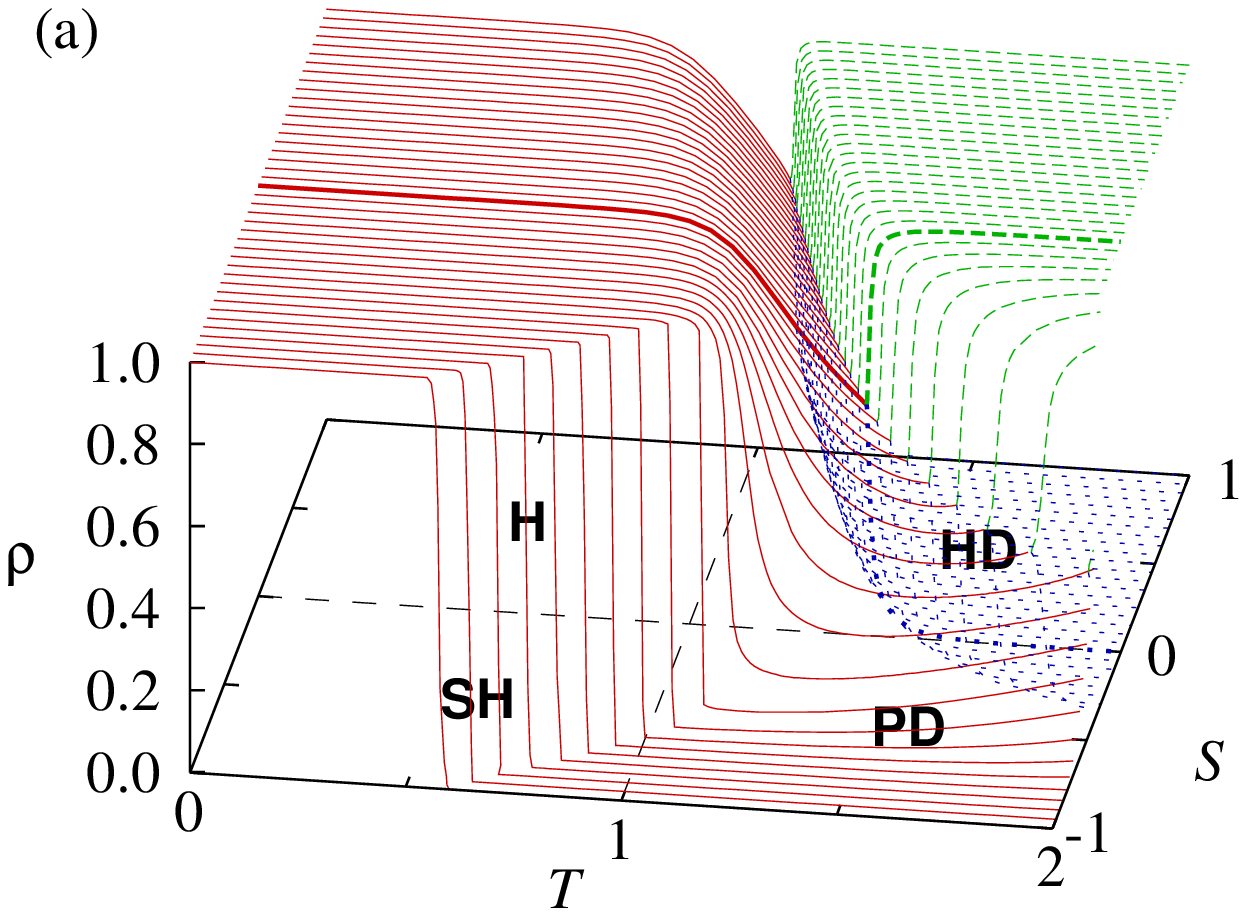,width=7cm}}
\centerline{\epsfig{file=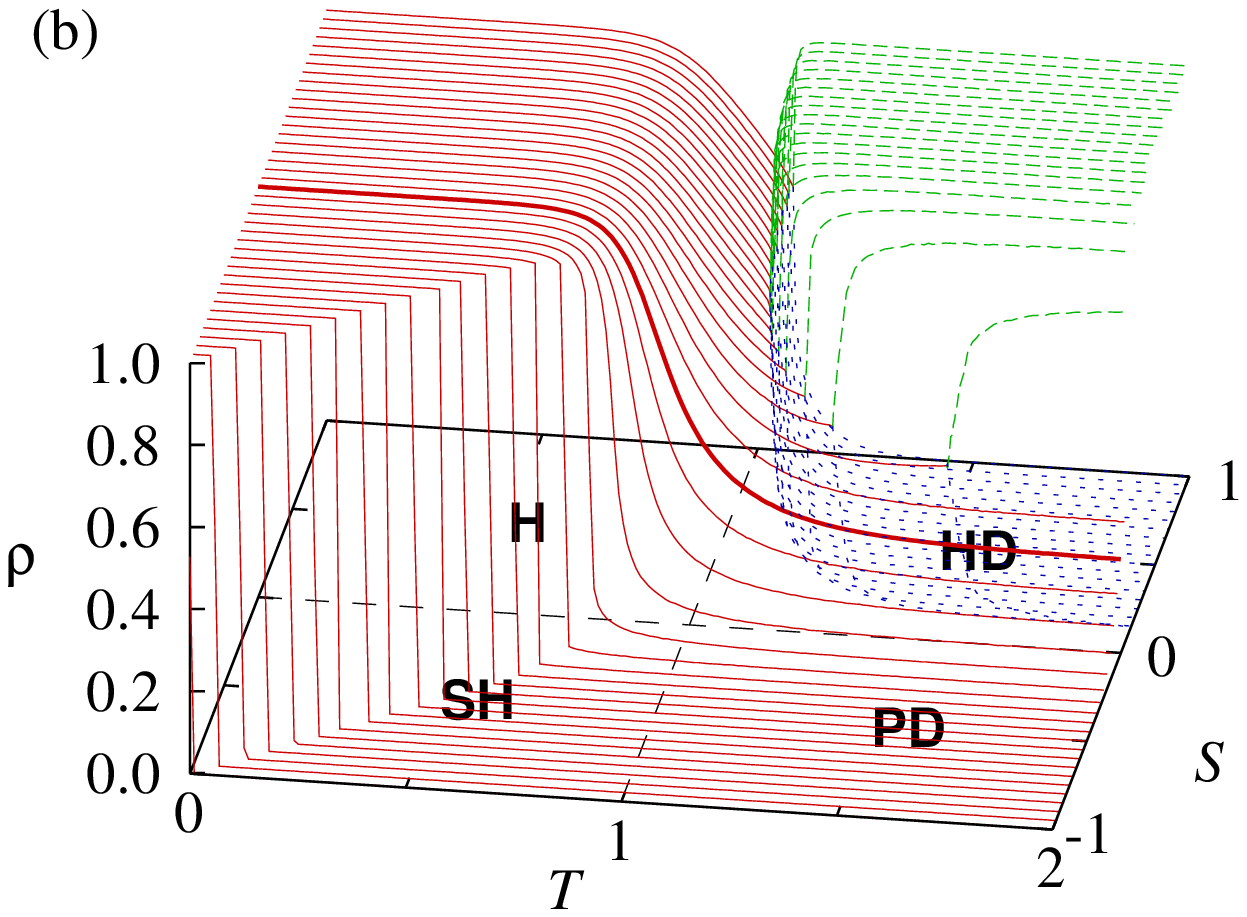,width=7cm}}
\centerline{\epsfig{file=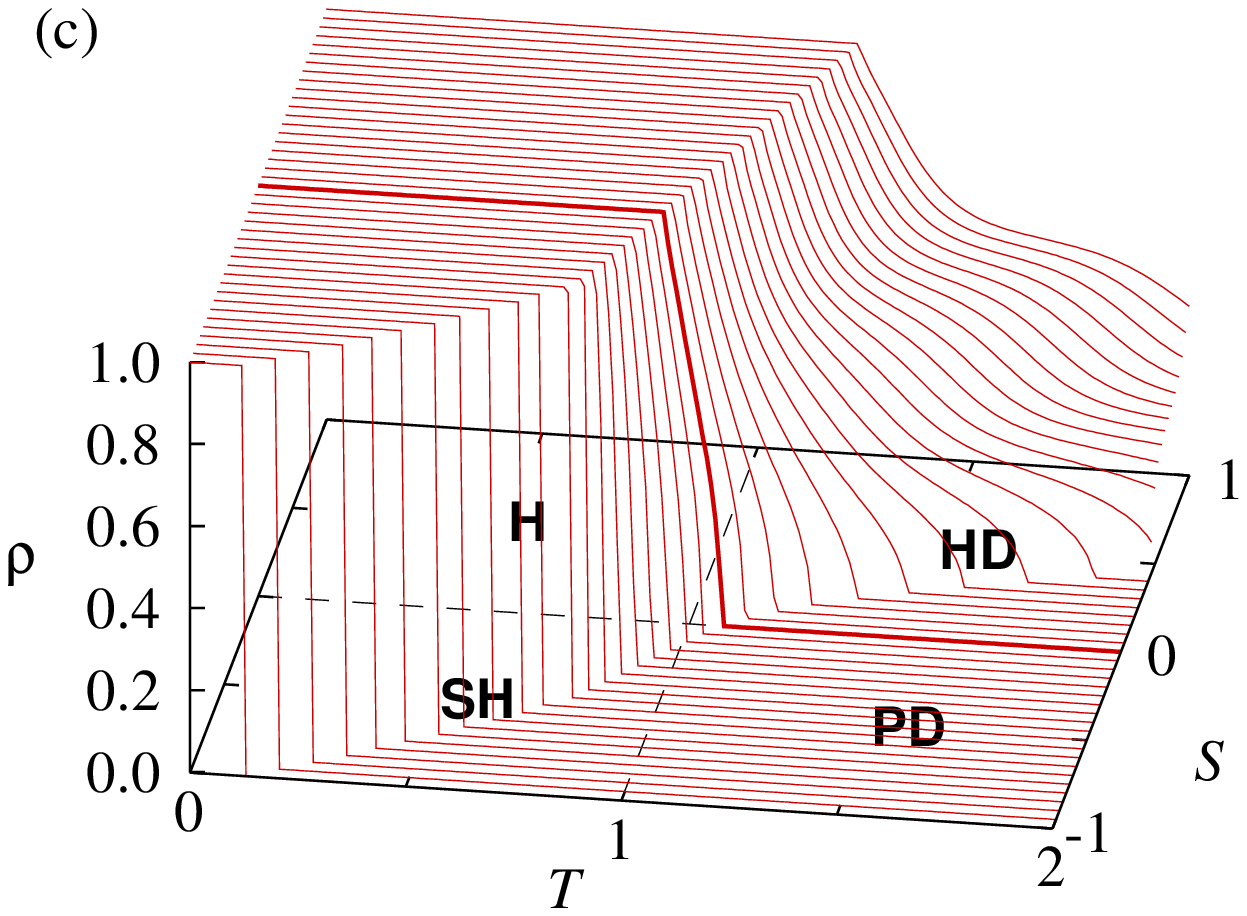,width=7cm}}
\caption{\label{fig:rho_ST}(color online) Fraction $\rho$ of cooperators in the two sublattices of the square lattice as a function of $T$ and $S$ for a fixed noise level ($K=0.25$). Plots from top to bottom illustrate the MC results when the evolution is controlled by pairwise collective (a), myopic (b), and pairwise imitative strategy updates (c). Dotted (blue) and dashed (green) lines denote distinguishable $\rho$s in the sublattices $A$ and $B$ while the solid (red) lines indicate $\rho$s in the absence of this symmetry breaking. The four quadrants of the $T-S$ plane correspond to games denoted by their abbreviation, namely, H="Harmony", HD="Hawk-Dove", SH="Stag Hunt", and PD="Prisoner's Dilemma". Thicker lines indicate data illustrated in the lower plot of Fig.~\ref{fig:rho_wPD}.}
\end{figure}

To reveal the possible sublattice ordering we carried out series of MC simulations for different $S$ values by using the same $K=0.25$ noise level. The results are summarized by consecutive curves in Fig.~\ref{fig:rho_ST} allowing us to compare the territories of $T-S$ parameters where cooperators, defectors, or sublattice ordering prevail the stationary state for the three dynamical rules we studied. It can be clearly seen, for example, that the defectors dominated area of $T-S$ region shrank significantly if the evolution is controlled by pairwise collective strategy update. For this dynamical rule players favor to choose cooperation in the sea of defectors if $T+4S>0$. This is the reason why the fraction of cooperators is sufficiently high in the case of weak PD. Notice furthermore that the pairwise collective strategy update provides the best condition for the cooperators to prevail the system within the region of SH game. Referring to the previously discussed connection between AC model and anti-ferromagnetic Ising model, SH game allows comparison with ferromagnetic ordering. The application of pairwise collective update reveals this possibility and highly extend the $C$-dominated phase in the SH quadrant.

The upper (a) plot of Fig.~\ref{fig:rho_ST} illustrates that sublattice ordering occurs over a large region of $T-S$ payoff parameters within the territory of HD and PD games if the evolution is controlled by pairwise collective strategy update. The critical temptation parameter $T_c(K)$ varies monotonously with the sucker's payoff $S$. 

For myopic strategy update the sublattice ordering can be observed only within the HD region of the payoff parameters as shown in the middle (b) plot of Fig.~\ref{fig:rho_ST}. It is emphasized that this behavior is predicted by the mean-field calculation for the case of weak PD, that is at the boundary separating the territories of HD and PD (see the lower plot of Fig.~\ref{fig:rho_wPD}). Comparing the $\rho(T,S)$ surfaces of Fig.~\ref{fig:rho_ST}a\&b, it is worth noting that myopic update does not support cooperation in the SH quadrant as effectively as it is done by pairwise collective update.

Finally, Fig.~\ref{fig:rho_ST}c indicates clearly that the imitation of the nearest neighbors on a square lattice does not support the emergence chessboard-like sublattice ordering. On the other hand, the upper and lower plateaus (at $\rho_A=\rho_B=0$ and 1) represent absorbing states and the continuous transitions to these homogeneous states belong to the directed percolation universality class \cite{szabo_pre98,chiappin_pre99}. Within the Stag Hunt region the system exhibits a first-order phase transition for all the three rules.

Evidently, the sublattice ordering is prevented if the noise level $K$ becomes sufficiently high for any values of $T$ and $S$. Simultaneously, the region of disordered coexistence of $C$ and $D$ strategies (where e.g., $0.01 < \rho_A$ and $\rho_B <0.99$) increases with $K$.

\section{Domain growth for sublattice ordering}
\label{sec:domgr}

The above simulations justified the appearance sublattice ordering for two of three dynamical rules we studied. It is emphasized, however, that many versions of similar spatial social dilemmas were investigated previously without reporting long-range ordering. Most of these investigations are based on imitation \cite{hauert_n04,roca_pre09,tomassini_pre06,roca_epjb09} preventing the formation of this type of ordered strategy distribution as mentioned above. The sublattice ordering is observable within small domains in the snapshots published previously by several authors \cite{sysiaho_epjb05,wang_pre06}. Sysi-Aho {\it et al.} studied HD game with myopic agents on a square lattice with first- and second-neighbor interactions \cite{sysiaho_epjb05}. In the latter model the randomly chosen agents are allowed to modify their strategy only if this action increases their own payoff (in contrary to the present stochastic myopic rule (Eq.~\ref{eq:myop}) allowing players to use of unfavored strategy with a low probability). Similar (short-range ordered) patterns were reported by Wang {\it et al.} \cite{wang_pre06} who used a more complicated synchronized strategy update.

\begin{figure}[ht]
\centerline{\epsfig{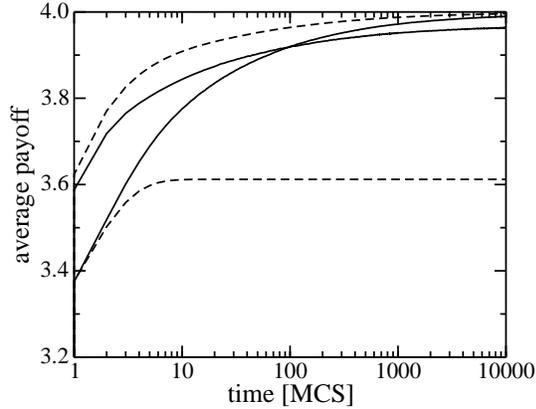}}
\caption{\label{fig:ut}The upper and lower solid lines illustrate the time dependence of the average payoff for the myopic (for $K=0.25$) and pairwise collective strategy adoption rules (for $K=0.25$) at $T=1.5$, $S=0.5$, and $L=4000$. The dashed lines denote the time dependence when the adoption of disadvantageous strategies is forbidden for the same parameters.}
\end{figure}

In order to clarify the importance of possible disadvantageous strategy change in the framework of myopic update, we compare the domain growth processes for two different cases. In the first case the evolution is controlled by the myopic rule as defined by Eq.~\ref{eq:myop}. In the second case we used the same strategy adoption probability (\ref{eq:myop}) but only if the new strategy increases the player's income [$P_x < P_x^{\prime}$] otherwise the adoption of the new strategy $s_x^{\prime}$ is forbidden. In other words, the second rule can be considered as a restriction of the first one when strategy change with payoff increment is allowed only. The visualization of the time-dependent strategy distribution indicated clearly that after a short ordering process the evolution is ended in a frozen pattern in the (restricted) second case if the simulation is started from a random initial state within the region of HD game. The difference between the time evolutions can be demonstrated if we compare the time-dependence of average payoffs $U(t)$ for both cases. As Figure~\ref{fig:ut} shows, the evolution of $U(t)$ stops very early (at about 15 MC steps) in the second case. In the first case, however, when disadvantageous strategy change is also allowed, $U(t)$ evolves continuously and saturates at $U_s=2(T+S)=4$ that corresponds to a mono-domain (long-range ordered) state. 

Regarding the restricted feature of the second update rule, a frozen pattern of strategy distribution can be considered as a state where every player is satisfied with her own strategy. In other words, a frozen state is analogous to a Nash equilibrium in the sense that the unilateral deviation from this strategy profile would reduce the income of the given player. As we demonstrated, a frozen state, that is the composition of two types of (equivalent) small ordered domains, can be avoided by applying the first rule because irrational strategy change along the interfaces helps to find global optimum. 

The above discussed phenomenon has inspired us to investigate what happens for the pairwise collective strategy update (defined by Eq.\ref{eq:pwc}) if the disadvantageous strategy adoptions are prohibited as earlier for myopic update. Surprisingly, the restriction of pairwise collective update does not block the domain growth process. The visualization of the evolution of strategy distribution indicates 
that strategies are not changed in the bulk of ordered domains but vary exclusively along the boundaries separating the ordered regions. Consequently, as Fig.~\ref{fig:ut} illustrates, $U(t)$ can increase continuously until reaching the value $U_s=4$ whether the strategy change is restricted to payoff increment cases or not.

\begin{figure}[ht]
\centerline{\epsfig{file=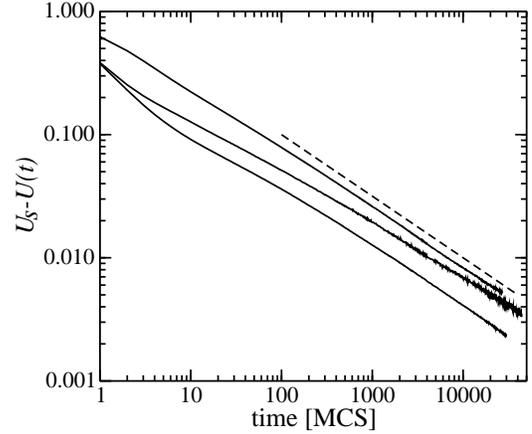,width=7cm}}
\caption{\label{fig:uta}Log-log plot of $U_s-U(t)$ versus time $t$ for (full) myopic, restricted, and full pairwise collective strategy adoption rules (from top to bottom). The MC data and parameters are the same plotted in Fig.~\ref{fig:ut}. The dashed line denotes the slope ($-1/2$) of theoretical prediction.}
\end{figure}

During the domain growth process the deficiency of average payoff is located along the boundaries separating the ordered regions therefore $U_s-U(t)$ decreases proportionally to the total length of interfaces.
This behavior resembles to the domain growth of solid-state physics systems where the growth kinetics is driven by reducing interfacial energy. In the latter, so-called curvature-driven growth the excess domain-boundary energy decays algebraically with $n=1/2$ exponent \cite{bray_ap94}. The time evolution of payoff difference in Fig.~\ref{fig:uta} supports our argument and shows algebraic decay with the mentioned exponent whether myopic or pairwise collective update was applied.

Figure \ref{fig:uta} illustrates that the deficiency of average payoffs vanishes algebraically, that is, $U_s-U(t) \simeq \Gamma / \sqrt{t}$ for all the three cases exhibiting growing domains. The speed (pre-factor $\Gamma$) of ordering decreases fast with the noise level $K$ because the disadvantageous strategy updates become rarer and rarer for sufficiently low values of noise. In the opposite case, when $K$ approaches its critical value $K_c$ dependent on the payoff parameters (ordering exists only if $K <K_c$), the domain growth also slows down due to the diverging fluctuations. In the quantitative comparison of domain growth for different dynamical rules the mentioned effects are compensated by doubling the value of $K$ for the pairwise collective strategy update  plotted in Figs. \ref{fig:ut} and \ref{fig:uta}. Surprisingly, for the latter dynamical rules the blocking of the disadvantageous strategy adoptions makes the domain growth faster (this phenomenon might have been related to the reduced noise effects along the interfaces).

Finally we mention that the preliminary simulations have confirmed the appearance of sublattice ordering for the models with nearest- and next-nearest interactions. More precisely, in the latter case one can observe a four-sublattice ordering process with many different types of ordered structures resembling those periodic structures described by a two-dimensional Ising model with first- and second-neighbor interactions \cite{binder_prb80,yin_pre09}. The poly-domain versions of most of these ordered structures were reported previously by several authors who studied the evolutionary HD games with a myopic strategy update when prohibiting the acceptance of those strategies yielding a lower individual payoff \cite{sysiaho_epjb05,roca_epjb09}. We have checked what happens if the present myopic evolutionary rule [given by Eq. (\ref{eq:myop})] is applied only if $(P_x-P_x^{\prime})<0$. In the latter case the domain growing process is also blocked in a (frozen) poly-domain state as it is described above.

\section{Conclusions and outlook}
\label{sec:conc}

In this work we have introduced a pairwise collective strategy update and studied its impact on anti-coordination and social dilemma games. Our proposal was motivated by real-life experiences when players act to increase not only personal but their local neighborhoods' payoff, too. As a starting effort along this avenue we have chosen pairs of players. To identify and quantify the consequence of this strategy update, we have also studied two frequently applied strategy updates, such as myopic and pairwise imitation. 

Our results highlight the emergence of a spatially ordered distribution of strategies on the analogy to anti-ferromagnetic ordering in spin systems. In contrary to the traditional imitations both the myopic and pairwise collective strategy updates support the formation of ordered strategy distribution favored within a wide range of social dilemmas. This ordered arrangement of cooperators and defectors can provide the maximum total payoff in a wide range of payoff parameters for the social dilemmas and it seems to be a general behavior on regular networks characterizing connections between the players. Within the context of social sciences the appearance of mentioned state can be interpreted as a possibility for the community to avoid the "tragedy of the common" (when all the agents choose $D$ and receive nothing) by sharing the two possible roles (strategies) with a spatially ordered (rigid) structure. The systematic comparison of the level of cooperations ($\rho$) for the three different evolutionary rules has justified that the pairwise collective strategy update provides the highest total (average) income for the whole spatial community in most of the region of payoff parameters. 

We have also studied the kinetics of ordering and found further similarities with physics motivated systems. This investigation highlighted the importance of irrational decisions as a way to avoid trapped (frozen) states.

There are two main directions to extend the present work. Firstly, as we already noted, other type of interaction graphs can also be considered. Our preliminary results confirm that the positive impact of the pairwise collective strategy update is not restricted to square lattice with nearest neighbor interaction. A similar ordering can emerge locally for a wide range of interaction graphs. Secondly, the size of the group in which players favor the group interest (instead of personal payoff) can be also increased. In this case further improvement of cooperation is expected.

\begin{acknowledgments}

This work was supported by the Hungarian National Research Fund
(Grant No. K-73449) and Bolyai Research Grant.

\end{acknowledgments}


\end{document}